\documentclass[sigconf]{acmart}

\usepackage{bm}

\AtBeginDocument{%
  \providecommand\BibTeX{{%
    \normalfont B\kern-0.5em{\scshape i\kern-0.25em b}\kern-0.8em\TeX}}}

\copyrightyear{2022}
\acmYear{2022}
\setcopyright{acmlicensed}
\acmConference[SIGIR '22] {Proceedings of the 45th International ACM SIGIR Conference on Research and Development in Information Retrieval}{July 11--15, 2022}{Madrid, Spain.}
\acmBooktitle{Proceedings of the 45th International ACM SIGIR Conference on Research and Development in Information Retrieval (SIGIR '22), July 11--15, 2022, Madrid, Spain}
\acmPrice{15.00}
\acmISBN{978-1-4503-8732-3/22/07}
\acmDOI{10.1145/3477495.3531861}
\begin{document}
\fancyhead{}

\title{Improving Micro-video Recommendation via Contrastive Multiple Interests}


\author{Beibei Li}
\affiliation{%
  \institution{Institute of Software, Chinese
Academy of Science}
  \institution{University of Chinese Academy of Sciences}
  \city{Beijing}
  \country{China}}
\email{libeibei16@otcaix.iscas.ac.cn}

\author{Beihong Jin}
\affiliation{%
  \institution{Institute of Software, Chinese
Academy of Science}
  \institution{University of Chinese Academy of Sciences}
  \city{Beijing}
  \country{China}}
\email{Beihong@iscas.ac.cn}

\author{Jiageng Song}
\affiliation{%
  \institution{Institute of Software, Chinese
Academy of Science}
  \institution{University of Chinese Academy of Sciences}
  \city{Beijing}
  \country{China}}
\email{songjiageng20@otcaix.iscas.ac.cn}

\author{Yisong Yu}
\affiliation{%
  \institution{Institute of Software, Chinese
Academy of Science}
  \institution{University of Chinese Academy of Sciences}
  \city{Beijing}
  \country{China}}
\email{yuyisong20@otcaix.iscas.ac.cn}

\author{Yiyuan Zheng}
\affiliation{%
  \institution{Institute of Software, Chinese
Academy of Science}
  \institution{University of Chinese Academy of Sciences}
  \city{Beijing}
  \country{China}}
\email{zhengyiyuan22@otcaix.iscas.ac.cn}

\author{Wei Zhuo}
\affiliation{%
  \institution{MX Media Co., Ltd.}
  \country{Singapore}}
\email{zhuowei@mx.in}








\begin{abstract}
With the rapid increase of micro-video creators and viewers, how to make personalized recommendations from a large number of candidates to viewers begins to attract more and more attention. However, existing micro-video recommendation models rely on expensive multi-modal information and learn an overall interest embedding that cannot reflect the user’s multiple interests in micro-videos. Recently, contrastive learning provides a new opportunity for refining the existing recommendation techniques. Therefore, in this paper, we propose to extract contrastive multi-interests and devise a micro-video recommendation model CMI. Specifically, CMI learns multiple interest embeddings for each user from his/her historical interaction sequence, in which the implicit orthogonal micro-video categories are used to decouple multiple user interests. Moreover, it establishes the contrastive multi-interest loss to improve the robustness of interest embeddings and the performance of recommendations. The results of experiments on two micro-video datasets demonstrate that CMI achieves state-of-the-art performance over existing baselines.
\end{abstract}


\ccsdesc[500]{Information systems~Recommender systems}

\keywords{Micro-video recommendation, Contrastive learning, Multi-interest learning}


\maketitle

\section{Introduction}

\begin{figure}[tb]
	\centering
	\includegraphics[height=2.3cm]{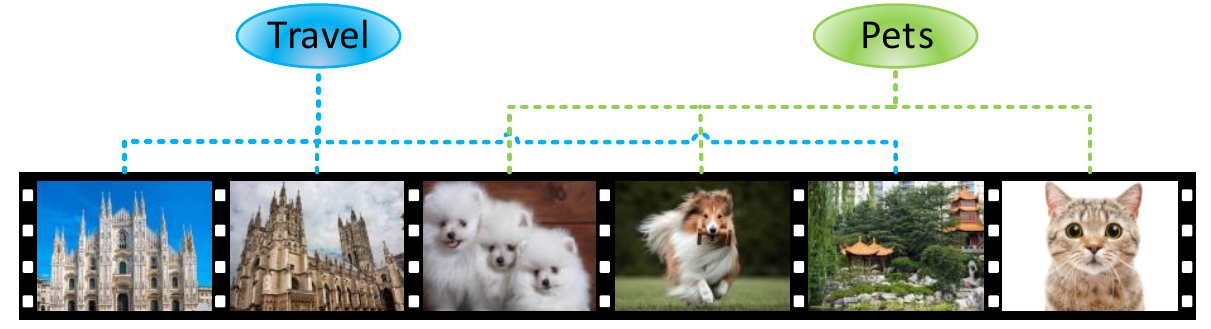}
	\caption{An micro-video interaction sequence with two interests: travel and pets.}
	\label{fig:ninterests}
\end{figure}

In recent years, micro-video apps such as TikTok, Kwai, MX TakaTak, etc. have become increasingly popular. Generally, the micro-video app displays a single video in full-screen mode at a time and automatically plays it in a repetitive way. Usually only after viewing the cover of the micro-video or watching the content of the micro-video for a few seconds, can the user determine whether he or she is interested in this micro-video. With the explosive growth of the number of micro-videos, if the micro-videos exposed to a user do not fall within the scope of his/her interests, then the user might leave the app. Therefore, the efficient micro-video recommendation has become a crucial task.

Existing micro-video recommendation models \cite{wei_mmgcn_2019,chen_temporal_2018,liu_user-video_2019,jiang_what_2020} rely on multi-modal information processing, which is too expensive to deal with large-scale micro-videos. Furthermore, they learn a single interest embedding for a user from his/her interaction sequence. However, most users have multiple interests while watching micro-videos. As shown in Figure \ref{fig:ninterests}, a user is interested in both tourism and pets, and the user's interactions in the future might involve any one of the user interests. Therefore, a more reasonable approach is to learn multiple disentangled interest embeddings for users, each of which represents one aspect of user interests, and then generate recommendations for the users based on the learned multiple disentangled interest embeddings.

On the other hand, contrastive learning has attracted a great deal of attention recently. It augments data to discover the implicit supervision signals in the input data and maximize the agreement between differently augmented views of the same data in a certain latent space. It has obtained the success in computer vision \cite{chen_simple_2020}, natural language processing \cite{fang_cert_2020,wu_clear_2020,giorgi_declutr_2021} and other domains \cite{oord_representation_2019}. More recently, contrastive learning also has been introduced to the recommendation, such as sequential recommendation, recommendation based on graph neural network, and etc., which realizes the debiasing \cite{zhou_contrastive_2021} and the denoising \cite{qin_world_2021}, and resolves the representation degeneration \cite{qiu_contrastive_2021} and the cold start problem \cite{wei_contrastive_2021}, improving the recommendation accuracy \cite{liu_contrastive_2021,xie_contrastive_2021,wu_self-supervised_2021,yu_graph_2021}. We note that there exists noise in the positive interactions in the micro-video scenario since micro-videos are automatically played and sometimes users cannot judge whether they like the micro-video or not until the micro-video finishes playing. However, neither existing micro-video recommendation models nor multi-interest recommendation models \cite{ma_disentangled_2020,cen_controllable_2020,liu_octopus_2020,li_multi-interest_2019} utilize contrastive learning to reduce the impact of noise in the positive interactions.

In this paper, we propose a new micro-video recommendation model named CMI. Based on the implicit micro-video category information, this model learns multiple disentangled interests for a user from his/her historical interaction sequence, recalls a group of micro-videos by each interest embedding, and then forms the final recommendation result. In particular, contrastive learning is incorporated into CMI to realize  the positive interaction denoising, improving the robustness of multi-interest disentanglement. Our contribution is summarized as follows.

\begin{enumerate}
    \item We propose CMI, a micro-video recommendation model, to explore the feasibility of combining contrastive learning with the multi-interest recommendation.
    \item We establish a multi-interest encoder based on implicit categories of items, and propose a contrastive multi-interest loss to minimize the difference between interests extracting from two augmented views of the same interaction sequence.
    \item We conduct experiments on two micro-video datasets and the experiment results show the effectiveness and rationality of the model. 
    
\end{enumerate}

\section{Methodology}
\begin{figure}[tb]
	\centering
	\includegraphics[height=4.9cm]{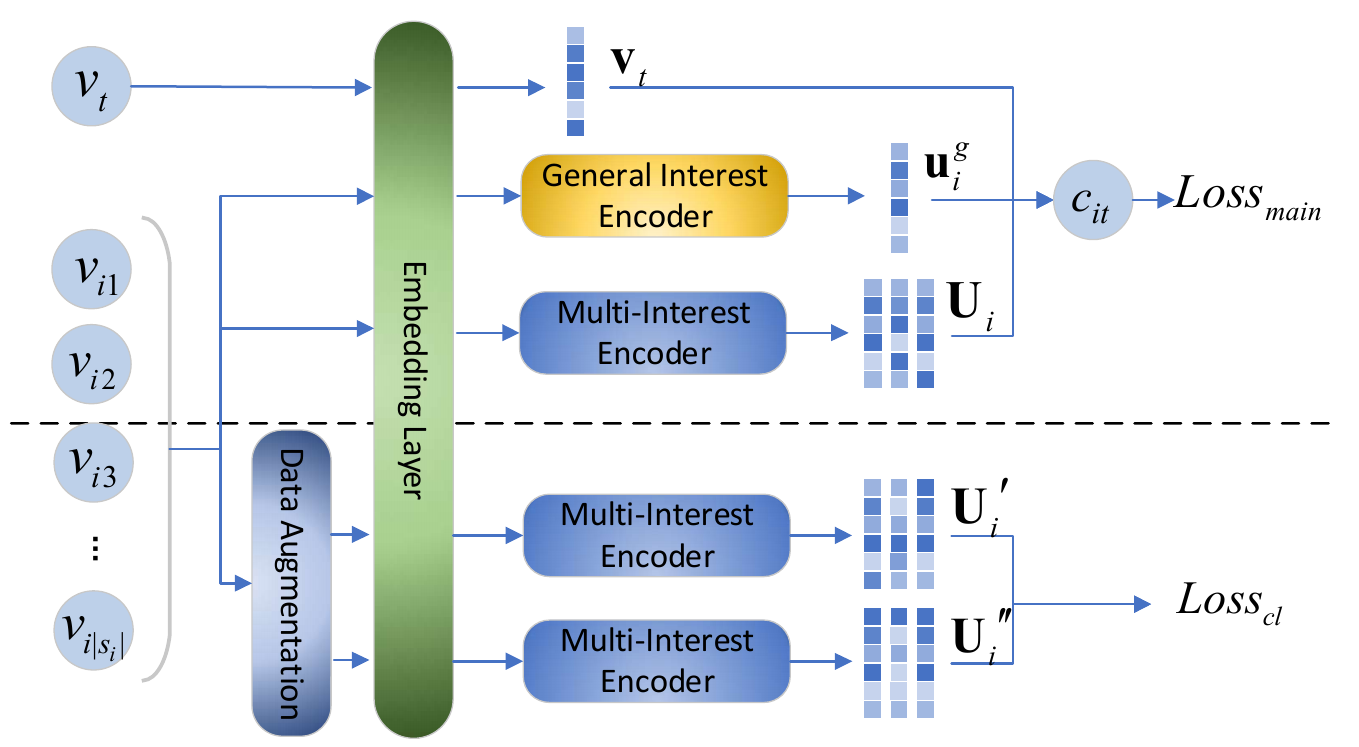}
	\caption{The architecture of CMI}
	\label{fig:model_fig}
\end{figure}

We denote user and item sets as $\mathcal{U}$ and $\mathcal{V}$, respectively. Further, we denote an interaction between a user and an item as a triad. That is, the fact that user $u_{i}$ interacts with micro-video $v_{j}$ at timestamp $t$ will be represented by $\left(i, j, t\right)$. 

Given a specific user $u_{i} \in \mathcal{U}$, we firstly generate a historical interaction sequence over a period of time, denoted as $s_{i}=\left[v_{i 1}, v_{i 2}, \ldots, v_{i\left|s_{i}\right|}\right]$, in which the videos are sorted by the timestamp that user $u_{i}$ interacts with the video in ascending order, and secondly learn multiple interest embeddings for each user, denoted as $\left[\mathbf{u}_{i}^{1}, \mathbf{u}_{i}^{2}, \ldots, \mathbf{u}_{i}^{m}\right]$. Then, for each interest embedding, we calculate the cosine similarity to each candidate micro-videos and recall $K$ micro-videos with the highest $K$ similarities, that is, a total of $m K$ micro-videos are recalled. Finally, from the recalled micro-videos, we select top-$K$ ones ranked by the cosine similarity as the final recommendations.

\subsection{Multi-interest and General Interest Encoders}

We argue that the categories of items are the basis of user interests. User preferences on a certain category of items form an interest of the user. Thus, we assume there are $m$ global categories and set learnable implicit embeddings $\left[\mathbf{g}_{1}, \mathbf{g}_{2}, \ldots, \mathbf{g}_{m}\right]$ for these $m$ categories. For items in historical interaction sequence $s_{i}$ of user $u_{i}$, we obtain the embedding of each item sequentially through the embedding layer, and form $\mathbf{S}_{i}=$ $\left[\mathbf{v}_{i 1}, \mathbf{v}_{i 2}, \ldots, \mathbf{v}_{i\left|s_{i}\right|}\right]$. We use the cosine similarity between an item embedding and a category embedding as the score that measures how the item belongs to the category. More specifically, the score of item $v_{i k} \in s_{i}$ matching category $l$ is calculated by Equation \ref{eq:wikj}.

\begin{equation}
\label{eq:wikj}
w_{i k}^{l}=\frac{\mathbf{g}_{l}^{T} \mathbf{v}_{i k}}{\left\|\mathbf{g}_{l}\right\|_{2}\left\|\mathbf{v}_{i k}\right\|_{2}}
\end{equation}

Next, the probability of item $v_{i k} \in s_{i}$ assigned to category $l$ is calculated by Equation \ref{eq:pikl}, where $\epsilon$ is a hyper-parameter smaller than 1 to avoid over-smoothing of probabilities.

\begin{equation}
\label{eq:pikl}
p_{i k}^{l}=\frac{\exp \left(w_{i k}^{l} / \epsilon\right)}{\sum_{l=1}^{m} \exp \left(w_{i k}^{l} / \epsilon\right)}    
\end{equation}

Then, the user interest $\mathbf{u}_{i}^{l}$ corresponding to the item category ${l}$ is calculated by Equation \ref{eq:uil}.

\begin{equation}
\label{eq:uil}
\mathbf{u}_{i}^{l}=\Sigma_{k=1}^{\left|s_{i}\right|} p_{i k}^{l} \mathbf{v}_{i k}  
\end{equation}


While performing the category assignment, we might encounter two degeneration cases. One is that each item has the same or similar probability of belonging to different categories. The reason of causing this degeneration is that learned item category embeddings are quite same with each other. The other is that one item category dominates the entire item embedding space, which means all items belong to that category. In order to avoid degeneration cases, we constrain both category embeddings and item embeddings within a unit hypersphere, that is,  $\lVert\mathbf{g}_{i}\rVert_{2}=\lVert \mathbf{v}_{*} \rVert_{2}=1$, and constrain every two category embeddings to be orthogonal, thus constructing an orthogonality loss as shown in Equation \ref{eq:lg}.

\begin{equation}
\label{eq:lg}
\mathcal{L}_{orth}=\sum_{i=1}^{m} \sum_{j=1, j \neq i}^{m} (\mathbf{g}_{i}^{T} \mathbf{g}_{j})^2
\end{equation}

In addition to encoding a user's multiple interests, we use GRU \cite{hidasi_recurrent_2018} to model the evolution of the general interest of the user, attaining the user's general  interest  $\mathbf{u}_{i}^{g}=G R U\left(\left[\mathbf{v}_{i1}, \mathbf{v}_{i2}, \ldots, \mathbf{v}_{i\left|s_{i}\right|}\right]\right)$. 

\subsection{Contrastive Regularization}

 We hold the view that user interests implied in the partial interactions are as same as ones implied in all the interactions. Therefore, we employ random sampling for data augmentation. Specifically, given the historical interaction sequence $s_i=[v_{i1},…,v_{i|s_i|}]$ of user $u_i$, we sample $\min \left(\mu\left|s_{i}\right|, f\right)$ micro-videos from $s_i$ and  form a new sequence $s_{i}^{\prime}$ according to their orders in $s_{i}$, where $\mu$ is the sampling ratio and $f$ is the longest sequence length whose default value is 100. By randomly sampling $s_{i}$ twice, we get two sequences $s_{i}^{\prime}$ and $s_{i}^{\prime \prime}$. Then we  feed these two augmented sequences to two multi-interest encoders to learn two groups of user interests, i.e.,  $\mathbf{U}_{i}^{\prime}=\left[\mathbf{u}_{i}^{1 \prime}, \mathbf{u}_{i}^{2 \prime}, \ldots, \mathbf{u}_{i}^{m \prime}\right]$ and $\mathbf{U}_{i}^{\prime \prime}=\left[\mathbf{u}_{i}^{1 \prime \prime}, \mathbf{u}_{i}^{2 \prime \prime}, \ldots, \mathbf{u}_{i}^{m \prime \prime}\right]$, as shown in Equation \ref{eq:uiuii}, where both $\mathbf{u}_{i}^{k \prime}$ and $\mathbf{u}_{i}^{k \prime \prime}$ are interests corresponding to the $k$-th micro-video category. 

\begin{equation}
\label{eq:uiuii}
\begin{aligned}
\mathbf{U}_{i}^{\prime} &=\operatorname{Multi-Interest-Encoder}\left(s_{i}^{\prime}\right) \\
\mathbf{U}_{i}^{\prime \prime} &=\operatorname{Multi-Interest-Encoder}\left(s_{i}^{\prime \prime}\right)
\end{aligned}
\end{equation}


Then, we construct a contrastive multi-interest loss as follows. For any interest embedding $\mathbf{u}_{i}^{k \prime} \in \mathbf{U}_{i}^{\prime}$ of user $u_{i}$, we construct a positive pair $(\mathbf{u}_{i}^{k \prime},\mathbf{u}_{i}^{k \prime \prime})$, construct $2m-2$ negative pairs using $\mathbf{u}_{i}^{k \prime}$ and the other $2m-2$ interest embeddings of user $u_{i}$, i.e., $\mathbf{u}_{i}^{h \prime} \in \mathbf{U}_{i}^{\prime}$ and $\mathbf{u}_{i}^{h \prime \prime} \in \mathbf{U}_{i}^{\prime \prime}$, where $h \in[1, m], h \neq k$. Since $m$ is usually not too large, the number of above negative pairs is limited. Therefore, given $\mathbf{u}_{i}^{k \prime}$, we take the interest embeddings of every other user in the same batch to build extra negative pairs. To sum up, let the training batch be $\mathcal{B}$ and the batch size be $|\mathcal{B}|$, for each positive pair, there are $2m(|\mathcal{B}|-1)+2m-2=2(m|\mathcal{B}|-1)$ negative pairs, which forms the negative set $\mathcal{S}^-$. Further, the contrastive multi-interest loss is defined in Equation \ref{eq:lcl}, where $\operatorname{sim}(\mathbf{a}, \mathbf{b})=\mathbf{a}^{T} \mathbf{b}/\left(\lVert\mathbf{a}\rVert_2\lVert\mathbf{b}\rVert_2\tau\right)$ and $\tau$ is a temperature parameter \cite{liu_contrastive_2021}.


\begin{equation}
\label{eq:lcl}
\begin{aligned}
 \mathcal{L}_{c l}\left(\mathbf{u}_{i}^{k \prime}, \mathbf{u}_{i}^{k \prime\prime}\right)=    &  -\log \frac{e^{\operatorname{sim}\left(\mathbf{u}_{i}^{k \prime}, \mathbf{u}_{i}^{k \prime \prime}\right)}}{e^{\operatorname{sim}\left(\mathbf{u}_{i}^{k \prime}, \mathbf{u}_{i}^{k \prime \prime}\right)}+\sum_{\mathbf{s}^- \in \mathcal{S}^-}e^{\operatorname{sim}\left(\mathbf{u}_{i}^{k \prime}, \mathbf{s}^{-}\right)}}\\
     &  -\log \frac{e^{\operatorname{sim}\left(\mathbf{u}_{i}^{k \prime}, \mathbf{u}_{i}^{k \prime \prime}\right)}}{e^{\operatorname{sim}\left(\mathbf{u}_{i}^{k \prime}, \mathbf{u}_{i}^{k \prime \prime}\right)}+\sum_{\mathbf{s}^- \in \mathcal{S}^-}e^{\operatorname{sim}\left(\mathbf{u}_{i}^{k \prime\prime}, \mathbf{s}^{-}\right)}}
\end{aligned}
\end{equation}


Through data augmentation and the contrastive multi-interest loss, user interest learning is no longer sensitive to a specific positive interaction, thereby reducing the impact of noisy positive interactions and realizing positive interaction denoising.

\subsection{Loss Function}

The interaction score between user $u_{i}$ and candidate item $v_{t}$ is predicted as $c_{i t}=\max _{0<k \leq m}\left(\left\{\mathbf{u}_{i}^{k T} \mathbf{v}_{t} / \epsilon\right\}\right)+\mathbf{u}_{i}^{g T} \mathbf{v}_{t}$, in which $k \in[1, m]$.

In the training process, for each positive sample $v_{p}^{i}$ of user $u_{i}$, we need to randomly sample $n$ micro-videos that have never been interacted with from the full micro-videos as negative samples. However, in order to avoid high sampling cost, given a positive sample, we only sample one negative sample, that is, $n$ is 1. Besides, we take the positive sampling items and negative sampling items of other users in the same batch as the negative samples, thus forming a negative sample set $\mathcal{N}$. We then adopt the following cross-entropy loss as the main part of the loss.

\begin{equation}
\label{eq:lmain}
\mathcal{L}_{\text {main}}\left(u_{i}, v_{p}^{i}\right)=-\ln \frac{\exp \left(c_{i p}\right)}{\sum_{v_{*} \in \left\{\mathcal{N} \cup v_{p}^{i}\right\}} \exp \left(c_{i *}\right)}    
\end{equation}

Finally, our loss function is shown in Equation \ref{eq:ltotal}, where $\lambda_{*}$ is the regularization coefficient.

\begin{equation}
\label{eq:ltotal}
\mathcal{L}=\mathcal{L}_{\text {main }}+\lambda_{c l} \mathcal{L}_{c l}+\lambda_{ {orth }} \mathcal{L}_{orth}  
\end{equation}

\section{Experiments}

\subsection{Experiment Setup}

\subsubsection{Datasets}

We conduct experiments on two micro-video datasets.

\begin{enumerate}
    \item \textbf{WeChat}. This is a public dataset released by WeChat Big Data Challenge 2021\footnote{https://algo.weixin.qq.com/problem-description}. This dataset contains user interactions on WeChat Channels, including explicit satisfaction interactions such as likes and favorites and implicit engagement interactions such as playing.
    \item \textbf{TakaTak}. This dataset is collected from TakaTak, a micro-video app for Indian users. The dataset contains interaction records of 50,000 anonymous users in four weeks.
\end{enumerate}

The statistics of the two datasets are shown in Table \ref{tab:dataset_stat}. For a dataset spanning $h$ day, we construct the training set with the interactions in the first $h-2$ days, the validation set with the interactions in the $h-1$-th day, and the test set with the interactions of the $h$-th day.

\begin{table}[t]
\begin{center}
\caption{The statistics of the two datasets. } \label{tab:dataset_stat}
\begin{tabular}{ccccc}
  \toprule
  Dataset & \#Users & \#Micro-videos & \#Interactions & Density  \\
  \midrule
  WeChat & 20000 & 77557 & 2666296 & 0.17\% \\
  TakaTak & 50000 &	157691	& 33863980	& 0.45\% \\
  \bottomrule
\end{tabular}
\end{center}
\end{table}

\subsubsection{Metrics}

Here, Recall@K and HitRate@K are used as metrics to evaluate the quality of the recommendations. 

\begin{table*}[tb]
\caption{Recommendation accuracy on two datasets. \#I. denotes the number of interests. The number in a bold type is the best performance in each column. The underlined number is the second best in each column.} \label{tab:perform-cmp}
\setlength\tabcolsep{4.2pt}
\begin{center}
\begin{tabular}{l|c|c c c | c c c|c|c c c | c c c}
\toprule
&  \multicolumn{7}{|c|}{\textbf{\small{WeChat}}}& \multicolumn{7}{|c}{\textbf{\small{TakaTak}}}\\
\cmidrule(lr){2-15}
&  & \multicolumn{3}{|c|}{\textbf{\small{Recall}}}& \multicolumn{3}{|c|}{\textbf{\small{HitRate}}}&  & \multicolumn{3}{|c|}{\textbf{\small{Recall}}}& \multicolumn{3}{|c}{\textbf{\small{HitRate}}}\\
\cmidrule(lr){2-15}
 & \textbf{\#I.} & \textbf{@10} & \textbf{@20} & \textbf{@50} & \textbf{@10} & \textbf{@20} & \textbf{@50} & \textbf{\#I.} & \textbf{@10} & \textbf{@20} & \textbf{@50} & \textbf{@10} & \textbf{@20} & \textbf{@50}\\
\midrule

Octopus&1	&0.0057	&0.0125	&0.0400	&0.0442	&0.0917	&0.2332	&1	&0.0076	&0.0160	&0.0447	&0.1457	&0.2533	&0.4393\\
MIND&1	&0.0296	&0.0521	&0.1025	&0.1774	&0.2791	&0.4514	&1	&0.0222	&0.0389	&\underline{0.0773}	&0.2139	&0.3263	&0.4977\\
ComiRec-DR&1	&0.0292	&0.0525	&0.1049	&0.1790	&0.2893	&0.4621	&1	&0.0226	&0.0392	&0.0769	&0.2345	&0.3427	&0.5144\\
ComiRec-SA&1	&0.0297	&0.0538	&0.1079	&0.1806	&0.2938	&0.4684	&1	&\underline{0.0239}	&\underline{0.0409}	&0.0752	&\underline{0.2567}	&0.3665	&0.5207\\
DSSRec&1  &  \underline{0.0327}&\underline{0.0578}&\underline{0.1161}&\underline{0.1971}&\underline{0.3064}&\underline{0.4854}&8&\textbf{0.0244}&0.0408&0.0749&0.2558&\underline{0.3704}&\underline{0.5259}\\

\midrule
CMI&8	&\textbf{0.0424}	&\textbf{0.0717}	&\textbf{0.1342} 	&\textbf{0.2436}	&\textbf{0.3612}	&\textbf{0.5292}	&8	&0.0210	&\textbf{0.0415}	&\textbf{0.0877}	&\textbf{0.2912}	&\textbf{0.4172}	&\textbf{0.5744}\\
\midrule
Improv.&/ &29.66\%&24.05\%&15.59\%&23.59\%&17.89\%&9.02\% & /&/&1.72\%&17.09\%&13.84\%&12.63\%&9.22\%\\
\bottomrule
\end{tabular}
\end{center}
\end{table*}

\subsubsection{Competitors}

We choose the following multi-interest recommendation models as competitors. 

\begin{enumerate}
    \item \textbf{Ocotopus} \cite{liu_octopus_2020}: It constructs an elastic archive network to extract diverse interests of users.
    \item \textbf{MIND} \cite{li_multi-interest_2019}: It adjusts the dynamic routing algorithm in the capsule network to extract multiple interests of users. 
    \item \textbf{ComiRec-DR} \cite{cen_controllable_2020}: It adopts the original dynamic routing algorithm of the capsule network to learn multiple user interests. 
    \item \textbf{ComiRec-SA} \cite{cen_controllable_2020}: It uses a multi-head attention mechanism to capture the multiple interests of users.
    \item \textbf{DSSRec} \cite{ma_disentangled_2020}: It disentangles multiple user intentions through self-supervised learning.
\end{enumerate}

To be fair, we do not compare with models that rely on multi-modal information.

\subsubsection{Implementation Details}

We implement our model with PyTorch, and initialize the parameters with the uniform distribution $U(-\frac{1}{\sqrt{d}}, \frac{1}{\sqrt{d}})$, where $d$ is the dimension of embeddings. We optimize the model through Adam. Hyper-parameters $\epsilon$, $\tau$ and $\lambda_{cl}$ are searched in [1, 0.1, 0.01, 0.001, 0.0001, 0.00001], and finally we set  $\epsilon=0.1$, $\tau=0.1$, and $\lambda_{cl}=0.01$. $\lambda_{orth}$ is searched in $[15,10,5,1,0.5]$ and finally we set it to 10. The sampling rate $\mu$ is set to 0.5.

For the sake of fairness, in all the experiments, we set the embedding dimension to 64 and the batch size to 1024. We stop training when Recall@50 on the validation set has not been improved in 5 consecutive epochs on the WeChat dataset and 10 consecutive epochs on the Takatak dataset. Besides, for MIND, ComiRec-DR, ComiRec-SA, and DSSRec, we use the open-source code released on Github \footnote{https://github.com/THUDM/ComiRec}$^,$\footnote{https://github.com/abinashsinha330/DSSRec}.

\subsection{Performance Comparison}

The experimental results of performance comparison are shown in Table \ref{tab:perform-cmp}, from which we have the following observations. 

(1) Multi-interest competitors except for CMI, with few exceptions, reach the best results while the number of interests is 1, indicating that these models cannot effectively capture multiple interests of a user in micro-videos.  
  
(2) The two dynamic routing-based models MIND and ComiRec-DR are not as good as ComiRec-SA and DSSRec. This is probably because both MIND and ComiRec-DR do not fully utilize the sequential relationship between historical interactions, but ComiRec-SA and DSSRec do. In addition, DSSRec adopts a novel seq2seq training strategy that leverages additional supervision signals, thus obtaining better performance.


(3) Octopus performs worst. That is probably because it aggressively routes every item into one interest exclusively at the beginning of training, which makes it easy to trap the parameters in a local optimum. 

(4) On two datasets, CMI far outperforms the competitors on most metrics, which demonstrates that CMI generates recommendations with both high accuracy and excellent coverage. The reason is that we avoid model degeneration by setting the category embeddings orthogonal and  extract multiple heterogeneous user interests. In addition, we achieve positive interaction denoising via contrastive learning, which improves the robustness of interest embeddings.

\subsection{Ablation Study}

While setting the number of interests to 8, we observe the performance of two model variants: CMI-CL and CMI-G, where the former is the CMI removing the contrastive multi-interest loss and the latter is the CMI without the general interest encoder. From the Table \ref{tab:ablation}, we find the model variants suffer severe declines in performance. This confirms the feasibility and effectiveness of contrastive learning for multi-interest recommendation, and  shows that whether the candidate item matches the general user preferences is also important.

\begin{table}[htbp]
\caption{Ablation study on WeChat. The values in parentheses are the percentages of decline relative to the original model.} \label{tab:ablation}
\begin{center}
\setlength\tabcolsep{4pt}
\begin{tabular}{l|l|c c c c}
\toprule
&	& CMI-CL		&CMI-G		&CMI\\
\midrule
&@10	&0.039(-8.02\%)	&0.0342(-19.34\%)	&\textbf{0.0424}\\
Recall&@20	&0.0665(-7.25\%)	&0.0589(-17.85\%)	&\textbf{0.0717}\\
&@50	&0.1285(-4.25\%)	&0.1165(-13.19\%)	&\textbf{0.1342}\\
\midrule
&@10	&0.2286(-6.16\%)	&0.2061(-15.39\%)	&\textbf{0.2436}\\
HitRate&@20	&0.3443(-4.68\%)	&0.3181(-11.93\%)	&\textbf{0.3612}\\
&@50	&0.5188(-1.93\%)	&0.4935(-6.71\%)	&\textbf{0.5290}\\
\bottomrule
\end{tabular}
\end{center}
\end{table}

\begin{table}[htbp]
\caption{The effect of the number of interests on WeChat. } \label{tab:eff-ninterest}
\begin{center}
\setlength\tabcolsep{4pt}
\begin{tabular}{l|c|c c c  c c}
\toprule
 & \#I. & 1& 2 & 4 & 8 & 16 \\
 \midrule
 &@10	&0.0303 & 0.0404	&0.0409	&\textbf{0.0428}	&0.0412\\
Recall&@20	&0.0530&0.0699	&0.0694	&\textbf{0.0718}	&0.0700\\
&@50	&0.1039&0.1343	&0.1333	&\textbf{0.1364}	&0.1314\\
\midrule
&@10	&0.1969&0.2383	&0.2384	&\textbf{0.2458}	&0.2390\\
HitRate&@20	&0.3012&0.3547	&0.3516	&\textbf{0.3587}	&0.3557\\
&@50	&0.4646&0.5330	&0.5271	&\textbf{0.5322}	&0.5238\\

\bottomrule
\end{tabular}
\end{center}
\end{table}

\subsection{Effect of the Number of Interests}

We set the number of interests to [1, 2, 4, 8, 16] successively and conduct experiments. The experimental results are shown in Table \ref{tab:eff-ninterest}. It can be seen that CMI reaches the best performance when the number of interests is 8 rather than 1. This confirms the necessity and effectiveness of extracting multiple interests in the micro-video recommendation scenarios.

\section{Conclusion}

This paper proposes a micro-video recommendation model CMI. The CMI model devises a multi-interest encoder and constructs a contrastive multi-interest loss to achieve positive interaction denoising and recommendation performance improvement. The performance of CMI on two micro-video datasets far exceeds other existing multi-interest models. The results of ablation study demonstrate that fusing contrastive learning into multi-interest extracting in micro-video recommendation is feasible and effective.

\begin{acks}
This work was supported by the National Natural Science Foundation of China under Grant No. 62072450 and the 2021 joint project with MX Media.
\end{acks}


\bibliographystyle{ACM-Reference-Format}
\bibliography{sample-base}


\begin{thebibliography}{22}


\ifx \showCODEN    \undefined \def \showCODEN     #1{\unskip}     \fi
\ifx \showDOI      \undefined \def \showDOI       #1{#1}\fi
\ifx \showISBNx    \undefined \def \showISBNx     #1{\unskip}     \fi
\ifx \showISBNxiii \undefined \def \showISBNxiii  #1{\unskip}     \fi
\ifx \showISSN     \undefined \def \showISSN      #1{\unskip}     \fi
\ifx \showLCCN     \undefined \def \showLCCN      #1{\unskip}     \fi
\ifx \shownote     \undefined \def \shownote      #1{#1}          \fi
\ifx \showarticletitle \undefined \def \showarticletitle #1{#1}   \fi
\ifx \showURL      \undefined \def \showURL       {\relax}        \fi
\providecommand\bibfield[2]{#2}
\providecommand\bibinfo[2]{#2}
\providecommand\natexlab[1]{#1}
\providecommand\showeprint[2][]{arXiv:#2}

\bibitem[\protect\citeauthoryear{Cen, Zhang, Zou, Zhou, Yang, and Tang}{Cen
  et~al\mbox{.}}{2020}]%
        {cen_controllable_2020}
\bibfield{author}{\bibinfo{person}{Yukuo Cen}, \bibinfo{person}{Jianwei Zhang},
  \bibinfo{person}{Xu Zou}, \bibinfo{person}{Chang Zhou},
  \bibinfo{person}{Hongxia Yang}, {and} \bibinfo{person}{Jie Tang}.}
  \bibinfo{year}{2020}\natexlab{}.
\newblock \showarticletitle{Controllable {Multi}-{Interest} {Framework} for
  {Recommendation}}. In \bibinfo{booktitle}{\emph{Proceedings of the 26th {ACM}
  {SIGKDD} {International} {Conference} on {Knowledge} {Discovery} \& {Data}
  {Mining}}}. \bibinfo{publisher}{ACM}, \bibinfo{pages}{2942--2951}.
\newblock
\showISBNx{978-1-4503-7998-4}
\urldef\tempurl%
\url{https://doi.org/10.1145/3394486.3403344}
\showDOI{\tempurl}


\bibitem[\protect\citeauthoryear{Chen, Kornblith, Norouzi, and Hinton}{Chen
  et~al\mbox{.}}{2020}]%
        {chen_simple_2020}
\bibfield{author}{\bibinfo{person}{Ting Chen}, \bibinfo{person}{Simon
  Kornblith}, \bibinfo{person}{Mohammad Norouzi}, {and}
  \bibinfo{person}{Geoffrey Hinton}.} \bibinfo{year}{2020}\natexlab{}.
\newblock \showarticletitle{A {Simple} {Framework} for {Contrastive} {Learning}
  of {Visual} {Representations}}.
\newblock \bibinfo{journal}{\emph{Proceedings of the 37th International
  Conference on Machine Learnin}}  \bibinfo{volume}{119} (\bibinfo{date}{June}
  \bibinfo{year}{2020}), \bibinfo{pages}{1597--1607}.
\newblock
\urldef\tempurl%
\url{http://proceedings.mlr.press/v119/chen20j.html}
\showURL{%
\tempurl}


\bibitem[\protect\citeauthoryear{Chen, Liu, Zha, Zhou, Xiong, and Li}{Chen
  et~al\mbox{.}}{2018}]%
        {chen_temporal_2018}
\bibfield{author}{\bibinfo{person}{Xusong Chen}, \bibinfo{person}{Dong Liu},
  \bibinfo{person}{Zheng-Jun Zha}, \bibinfo{person}{Wengang Zhou},
  \bibinfo{person}{Zhiwei Xiong}, {and} \bibinfo{person}{Yan Li}.}
  \bibinfo{year}{2018}\natexlab{}.
\newblock \showarticletitle{Temporal {Hierarchical} {Attention} at {Category}-
  and {Item}-{Level} for {Micro}-{Video} {Click}-{Through} {Prediction}}. In
  \bibinfo{booktitle}{\emph{2018 {ACM} {Multimedia} {Conference} on
  {Multimedia} {Conference} - {MM} '18}}. \bibinfo{publisher}{ACM Press},
  \bibinfo{pages}{1146--1153}.
\newblock
\showISBNx{978-1-4503-5665-7}
\urldef\tempurl%
\url{https://doi.org/10.1145/3240508.3240617}
\showDOI{\tempurl}


\bibitem[\protect\citeauthoryear{Fang, Wang, Zhou, Ding, and Xie}{Fang
  et~al\mbox{.}}{2020}]%
        {fang_cert_2020}
\bibfield{author}{\bibinfo{person}{Hongchao Fang}, \bibinfo{person}{Sicheng
  Wang}, \bibinfo{person}{Meng Zhou}, \bibinfo{person}{Jiayuan Ding}, {and}
  \bibinfo{person}{Pengtao Xie}.} \bibinfo{year}{2020}\natexlab{}.
\newblock \showarticletitle{{CERT}: {Contrastive} {Self}-supervised {Learning}
  for {Language} {Understanding}}.
\newblock \bibinfo{journal}{\emph{arXiv:2005.12766 [cs, stat]}}
  (\bibinfo{date}{June} \bibinfo{year}{2020}).
\newblock
\urldef\tempurl%
\url{http://arxiv.org/abs/2005.12766}
\showURL{%
\tempurl}
\newblock
\shownote{arXiv: 2005.12766.}


\bibitem[\protect\citeauthoryear{Giorgi, Nitski, Wang, and Bader}{Giorgi
  et~al\mbox{.}}{2021}]%
        {giorgi_declutr_2021}
\bibfield{author}{\bibinfo{person}{John Giorgi}, \bibinfo{person}{Osvald
  Nitski}, \bibinfo{person}{Bo Wang}, {and} \bibinfo{person}{Gary Bader}.}
  \bibinfo{year}{2021}\natexlab{}.
\newblock \showarticletitle{{DeCLUTR}: {Deep} {Contrastive} {Learning} for
  {Unsupervised} {Textual} {Representations}}.
\newblock \bibinfo{journal}{\emph{arXiv:2006.03659 [cs]}} (\bibinfo{date}{May}
  \bibinfo{year}{2021}).
\newblock
\urldef\tempurl%
\url{http://arxiv.org/abs/2006.03659}
\showURL{%
\tempurl}
\newblock
\shownote{arXiv: 2006.03659.}


\bibitem[\protect\citeauthoryear{Hidasi and Karatzoglou}{Hidasi and
  Karatzoglou}{2018}]%
        {hidasi_recurrent_2018}
\bibfield{author}{\bibinfo{person}{Balázs Hidasi} {and}
  \bibinfo{person}{Alexandros Karatzoglou}.} \bibinfo{year}{2018}\natexlab{}.
\newblock \showarticletitle{Recurrent {Neural} {Networks} with {Top}-k {Gains}
  for {Session}-based {Recommendations}}. In
  \bibinfo{booktitle}{\emph{Proceedings of the 27th {ACM} {International}
  {Conference} on {Information} and {Knowledge} {Management}}}.
  \bibinfo{publisher}{ACM}, \bibinfo{pages}{843--852}.
\newblock
\showISBNx{978-1-4503-6014-2}
\urldef\tempurl%
\url{https://doi.org/10.1145/3269206.3271761}
\showDOI{\tempurl}


\bibitem[\protect\citeauthoryear{Jiang, Wang, Wei, Gao, Wang, and Nie}{Jiang
  et~al\mbox{.}}{2020}]%
        {jiang_what_2020}
\bibfield{author}{\bibinfo{person}{Hao Jiang}, \bibinfo{person}{Wenjie Wang},
  \bibinfo{person}{Yinwei Wei}, \bibinfo{person}{Zan Gao},
  \bibinfo{person}{Yinglong Wang}, {and} \bibinfo{person}{Liqiang Nie}.}
  \bibinfo{year}{2020}\natexlab{}.
\newblock \showarticletitle{What {Aspect} {Do} {You} {Like}: {Multi}-scale
  {Time}-aware {User} {Interest} {Modeling} for {Micro}-video
  {Recommendation}}. In \bibinfo{booktitle}{\emph{Proceedings of the 28th {ACM}
  {International} {Conference} on {Multimedia}}}. \bibinfo{publisher}{ACM},
  \bibinfo{pages}{3487--3495}.
\newblock
\showISBNx{978-1-4503-7988-5}
\urldef\tempurl%
\url{https://doi.org/10.1145/3394171.3413653}
\showDOI{\tempurl}


\bibitem[\protect\citeauthoryear{Li, Liu, Wu, Xu, Zhao, Huang, Kang, Chen, Li,
  and Lee}{Li et~al\mbox{.}}{2019}]%
        {li_multi-interest_2019}
\bibfield{author}{\bibinfo{person}{Chao Li}, \bibinfo{person}{Zhiyuan Liu},
  \bibinfo{person}{Mengmeng Wu}, \bibinfo{person}{Yuchi Xu},
  \bibinfo{person}{Huan Zhao}, \bibinfo{person}{Pipei Huang},
  \bibinfo{person}{Guoliang Kang}, \bibinfo{person}{Qiwei Chen},
  \bibinfo{person}{Wei Li}, {and} \bibinfo{person}{Dik~Lun Lee}.}
  \bibinfo{year}{2019}\natexlab{}.
\newblock \showarticletitle{Multi-{Interest} {Network} with {Dynamic} {Routing}
  for {Recommendation} at {Tmall}}. In \bibinfo{booktitle}{\emph{Proceedings of
  the 28th {ACM} {International} {Conference} on {Information} and {Knowledge}
  {Management}}}. \bibinfo{publisher}{ACM}, \bibinfo{pages}{2615--2623}.
\newblock
\showISBNx{978-1-4503-6976-3}
\urldef\tempurl%
\url{https://doi.org/10.1145/3357384.3357814}
\showDOI{\tempurl}


\bibitem[\protect\citeauthoryear{Liu, Chen, Liu, and Hu}{Liu
  et~al\mbox{.}}{2019}]%
        {liu_user-video_2019}
\bibfield{author}{\bibinfo{person}{Shang Liu}, \bibinfo{person}{Zhenzhong
  Chen}, \bibinfo{person}{Hongyi Liu}, {and} \bibinfo{person}{Xinghai Hu}.}
  \bibinfo{year}{2019}\natexlab{}.
\newblock \showarticletitle{User-{Video} {Co}-{Attention} {Network} for
  {Personalized} {Micro}-video {Recommendation}}. In
  \bibinfo{booktitle}{\emph{The {World} {Wide} {Web} {Conference} on - {WWW}
  '19}}. \bibinfo{publisher}{ACM Press}, \bibinfo{pages}{3020--3026}.
\newblock
\showISBNx{978-1-4503-6674-8}
\urldef\tempurl%
\url{https://doi.org/10.1145/3308558.3313513}
\showDOI{\tempurl}


\bibitem[\protect\citeauthoryear{Liu, Chen, Li, Yu, McAuley, and Xiong}{Liu
  et~al\mbox{.}}{2021}]%
        {liu_contrastive_2021}
\bibfield{author}{\bibinfo{person}{Zhiwei Liu}, \bibinfo{person}{Yongjun Chen},
  \bibinfo{person}{Jia Li}, \bibinfo{person}{Philip~S. Yu},
  \bibinfo{person}{Julian McAuley}, {and} \bibinfo{person}{Caiming Xiong}.}
  \bibinfo{year}{2021}\natexlab{}.
\newblock \showarticletitle{Contrastive {Self}-supervised {Sequential}
  {Recommendation} with {Robust} {Augmentation}}.
\newblock \bibinfo{journal}{\emph{arXiv:2108.06479 [cs]}} (\bibinfo{date}{Aug.}
  \bibinfo{year}{2021}).
\newblock
\urldef\tempurl%
\url{http://arxiv.org/abs/2108.06479}
\showURL{%
\tempurl}
\newblock
\shownote{arXiv: 2108.06479.}


\bibitem[\protect\citeauthoryear{Liu, Lian, Yang, Lian, and Xie}{Liu
  et~al\mbox{.}}{2020}]%
        {liu_octopus_2020}
\bibfield{author}{\bibinfo{person}{Zheng Liu}, \bibinfo{person}{Jianxun Lian},
  \bibinfo{person}{Junhan Yang}, \bibinfo{person}{Defu Lian}, {and}
  \bibinfo{person}{Xing Xie}.} \bibinfo{year}{2020}\natexlab{}.
\newblock \showarticletitle{Octopus: {Comprehensive} and {Elastic} {User}
  {Representation} for the {Generation} of {Recommendation} {Candidates}}. In
  \bibinfo{booktitle}{\emph{Proceedings of the 43rd {International} {ACM}
  {SIGIR} {Conference} on {Research} and {Development} in {Information}
  {Retrieval}}}. \bibinfo{publisher}{ACM}, \bibinfo{pages}{289--298}.
\newblock
\showISBNx{978-1-4503-8016-4}
\urldef\tempurl%
\url{https://doi.org/10.1145/3397271.3401088}
\showDOI{\tempurl}


\bibitem[\protect\citeauthoryear{Ma, Zhou, Yang, Cui, Wang, and Zhu}{Ma
  et~al\mbox{.}}{2020}]%
        {ma_disentangled_2020}
\bibfield{author}{\bibinfo{person}{Jianxin Ma}, \bibinfo{person}{Chang Zhou},
  \bibinfo{person}{Hongxia Yang}, \bibinfo{person}{Peng Cui},
  \bibinfo{person}{Xin Wang}, {and} \bibinfo{person}{Wenwu Zhu}.}
  \bibinfo{year}{2020}\natexlab{}.
\newblock \showarticletitle{Disentangled {Self}-{Supervision} in {Sequential}
  {Recommenders}}. In \bibinfo{booktitle}{\emph{Proceedings of the 26th {ACM}
  {SIGKDD} {International} {Conference} on {Knowledge} {Discovery} \& {Data}
  {Mining}}}. \bibinfo{publisher}{ACM}, \bibinfo{pages}{483--491}.
\newblock
\showISBNx{978-1-4503-7998-4}
\urldef\tempurl%
\url{https://doi.org/10.1145/3394486.3403091}
\showDOI{\tempurl}


\bibitem[\protect\citeauthoryear{Oord, Li, and Vinyals}{Oord
  et~al\mbox{.}}{2019}]%
        {oord_representation_2019}
\bibfield{author}{\bibinfo{person}{Aaron van~den Oord}, \bibinfo{person}{Yazhe
  Li}, {and} \bibinfo{person}{Oriol Vinyals}.} \bibinfo{year}{2019}\natexlab{}.
\newblock \showarticletitle{Representation {Learning} with {Contrastive}
  {Predictive} {Coding}}.
\newblock \bibinfo{journal}{\emph{arXiv:1807.03748 [cs, stat]}}
  (\bibinfo{date}{Jan.} \bibinfo{year}{2019}).
\newblock
\urldef\tempurl%
\url{http://arxiv.org/abs/1807.03748}
\showURL{%
\tempurl}
\newblock
\shownote{arXiv: 1807.03748.}


\bibitem[\protect\citeauthoryear{Qin, Wang, and Li}{Qin et~al\mbox{.}}{2021}]%
        {qin_world_2021}
\bibfield{author}{\bibinfo{person}{Yuqi Qin}, \bibinfo{person}{Pengfei Wang},
  {and} \bibinfo{person}{Chenliang Li}.} \bibinfo{year}{2021}\natexlab{}.
\newblock \showarticletitle{The {World} is {Binary}: {Contrastive} {Learning}
  for {Denoising} {Next} {Basket} {Recommendation}}. In
  \bibinfo{booktitle}{\emph{Proceedings of the 44th {International} {ACM}
  {SIGIR} {Conference} on {Research} and {Development} in {Information}
  {Retrieval}}}. \bibinfo{publisher}{ACM}, \bibinfo{pages}{859--868}.
\newblock
\showISBNx{978-1-4503-8037-9}
\urldef\tempurl%
\url{https://doi.org/10.1145/3404835.3462836}
\showDOI{\tempurl}


\bibitem[\protect\citeauthoryear{Qiu, Huang, Yin, and Wang}{Qiu
  et~al\mbox{.}}{2021}]%
        {qiu_contrastive_2021}
\bibfield{author}{\bibinfo{person}{Ruihong Qiu}, \bibinfo{person}{Zi Huang},
  \bibinfo{person}{Hongzhi Yin}, {and} \bibinfo{person}{Zijian Wang}.}
  \bibinfo{year}{2021}\natexlab{}.
\newblock \showarticletitle{Contrastive {Learning} for {Representation}
  {Degeneration} {Problem} in {Sequential} {Recommendation}}.
\newblock \bibinfo{journal}{\emph{arXiv:2110.05730 [cs]}} (\bibinfo{date}{Nov.}
  \bibinfo{year}{2021}).
\newblock
\urldef\tempurl%
\url{https://doi.org/10.1145/3488560.3498433}
\showDOI{\tempurl}
\newblock
\shownote{arXiv: 2110.05730.}


\bibitem[\protect\citeauthoryear{Wei, Wang, Li, Nie, Li, Li, and Chua}{Wei
  et~al\mbox{.}}{2021}]%
        {wei_contrastive_2021}
\bibfield{author}{\bibinfo{person}{Yinwei Wei}, \bibinfo{person}{Xiang Wang},
  \bibinfo{person}{Qi Li}, \bibinfo{person}{Liqiang Nie}, \bibinfo{person}{Yan
  Li}, \bibinfo{person}{Xuanping Li}, {and} \bibinfo{person}{Tat-Seng Chua}.}
  \bibinfo{year}{2021}\natexlab{}.
\newblock \showarticletitle{Contrastive {Learning} for {Cold}-{Start}
  {Recommendation}}. In \bibinfo{booktitle}{\emph{Proceedings of the 29th {ACM}
  {International} {Conference} on {Multimedia}}}. \bibinfo{publisher}{ACM},
  \bibinfo{pages}{5382--5390}.
\newblock
\showISBNx{978-1-4503-8651-7}
\urldef\tempurl%
\url{https://doi.org/10.1145/3474085.3475665}
\showDOI{\tempurl}


\bibitem[\protect\citeauthoryear{Wei, Wang, Nie, He, Hong, and Chua}{Wei
  et~al\mbox{.}}{2019}]%
        {wei_mmgcn_2019}
\bibfield{author}{\bibinfo{person}{Yinwei Wei}, \bibinfo{person}{Xiang Wang},
  \bibinfo{person}{Liqiang Nie}, \bibinfo{person}{Xiangnan He},
  \bibinfo{person}{Richang Hong}, {and} \bibinfo{person}{Tat-Seng Chua}.}
  \bibinfo{year}{2019}\natexlab{}.
\newblock \showarticletitle{{MMGCN}: {Multi}-modal {Graph} {Convolution}
  {Network} for {Personalized} {Recommendation} of {Micro}-video}. In
  \bibinfo{booktitle}{\emph{Proceedings of the 27th {ACM} {International}
  {Conference} on {Multimedia}}}. \bibinfo{publisher}{ACM},
  \bibinfo{pages}{1437--1445}.
\newblock
\showISBNx{978-1-4503-6889-6}
\urldef\tempurl%
\url{https://doi.org/10.1145/3343031.3351034}
\showDOI{\tempurl}


\bibitem[\protect\citeauthoryear{Wu, Wang, Feng, He, Chen, Lian, and Xie}{Wu
  et~al\mbox{.}}{2021}]%
        {wu_self-supervised_2021}
\bibfield{author}{\bibinfo{person}{Jiancan Wu}, \bibinfo{person}{Xiang Wang},
  \bibinfo{person}{Fuli Feng}, \bibinfo{person}{Xiangnan He},
  \bibinfo{person}{Liang Chen}, \bibinfo{person}{Jianxun Lian}, {and}
  \bibinfo{person}{Xing Xie}.} \bibinfo{year}{2021}\natexlab{}.
\newblock \showarticletitle{Self-supervised {Graph} {Learning} for
  {Recommendation}}.
\newblock \bibinfo{journal}{\emph{arXiv:2010.10783 [cs]}} (\bibinfo{date}{May}
  \bibinfo{year}{2021}).
\newblock
\urldef\tempurl%
\url{https://doi.org/10.1145/3404835.3462862}
\showDOI{\tempurl}
\newblock
\shownote{arXiv: 2010.10783.}


\bibitem[\protect\citeauthoryear{Wu, Wang, Gu, Khabsa, Sun, and Ma}{Wu
  et~al\mbox{.}}{2020}]%
        {wu_clear_2020}
\bibfield{author}{\bibinfo{person}{Zhuofeng Wu}, \bibinfo{person}{Sinong Wang},
  \bibinfo{person}{Jiatao Gu}, \bibinfo{person}{Madian Khabsa},
  \bibinfo{person}{Fei Sun}, {and} \bibinfo{person}{Hao Ma}.}
  \bibinfo{year}{2020}\natexlab{}.
\newblock \showarticletitle{{CLEAR}: {Contrastive} {Learning} for {Sentence}
  {Representation}}.
\newblock \bibinfo{journal}{\emph{arXiv:2012.15466 [cs]}} (\bibinfo{date}{Dec.}
  \bibinfo{year}{2020}).
\newblock
\urldef\tempurl%
\url{http://arxiv.org/abs/2012.15466}
\showURL{%
\tempurl}
\newblock
\shownote{arXiv: 2012.15466.}


\bibitem[\protect\citeauthoryear{Xie, Sun, Liu, Wu, Gao, Ding, and Cui}{Xie
  et~al\mbox{.}}{2021}]%
        {xie_contrastive_2021}
\bibfield{author}{\bibinfo{person}{Xu Xie}, \bibinfo{person}{Fei Sun},
  \bibinfo{person}{Zhaoyang Liu}, \bibinfo{person}{Shiwen Wu},
  \bibinfo{person}{Jinyang Gao}, \bibinfo{person}{Bolin Ding}, {and}
  \bibinfo{person}{Bin Cui}.} \bibinfo{year}{2021}\natexlab{}.
\newblock \showarticletitle{Contrastive {Learning} for {Sequential}
  {Recommendation}}.
\newblock \bibinfo{journal}{\emph{arXiv:2010.14395 [cs]}} (\bibinfo{date}{Feb.}
  \bibinfo{year}{2021}).
\newblock
\urldef\tempurl%
\url{http://arxiv.org/abs/2010.14395}
\showURL{%
\tempurl}
\newblock
\shownote{arXiv: 2010.14395.}


\bibitem[\protect\citeauthoryear{Yu, Yin, Xia, Cui, and Nguyen}{Yu
  et~al\mbox{.}}{2021}]%
        {yu_graph_2021}
\bibfield{author}{\bibinfo{person}{Junliang Yu}, \bibinfo{person}{Hongzhi Yin},
  \bibinfo{person}{Xin Xia}, \bibinfo{person}{Lizhen Cui}, {and}
  \bibinfo{person}{Quoc Viet~Hung Nguyen}.} \bibinfo{year}{2021}\natexlab{}.
\newblock \showarticletitle{Graph {Augmentation}-{Free} {Contrastive}
  {Learning} for {Recommendation}}.
\newblock \bibinfo{journal}{\emph{arXiv:2112.08679 [cs]}} (\bibinfo{date}{Dec.}
  \bibinfo{year}{2021}).
\newblock
\urldef\tempurl%
\url{http://arxiv.org/abs/2112.08679}
\showURL{%
\tempurl}
\newblock
\shownote{arXiv: 2112.08679.}


\bibitem[\protect\citeauthoryear{Zhou, Ma, Zhang, Zhou, and Yang}{Zhou
  et~al\mbox{.}}{2021}]%
        {zhou_contrastive_2021}
\bibfield{author}{\bibinfo{person}{Chang Zhou}, \bibinfo{person}{Jianxin Ma},
  \bibinfo{person}{Jianwei Zhang}, \bibinfo{person}{Jingren Zhou}, {and}
  \bibinfo{person}{Hongxia Yang}.} \bibinfo{year}{2021}\natexlab{}.
\newblock \showarticletitle{Contrastive {Learning} for {Debiased} {Candidate}
  {Generation} in {Large}-{Scale} {Recommender} {Systems}}. In
  \bibinfo{booktitle}{\emph{Proceedings of the 27th {ACM} {SIGKDD} {Conference}
  on {Knowledge} {Discovery} \& {Data} {Mining}}}. \bibinfo{publisher}{ACM},
  \bibinfo{pages}{3985--3995}.
\newblock
\showISBNx{978-1-4503-8332-5}
\urldef\tempurl%
\url{https://doi.org/10.1145/3447548.3467102}
\showDOI{\tempurl}


\end{thebibliography}










\end{document}